\magnification=\magstep1
\baselineskip=17pt
\hfuzz=6pt

$ $

\vskip 1cm

\centerline{\bf A Turing test for free will}

\bigskip

\centerline{Seth Lloyd}

\centerline{Massachusetts Insitute of Technology}

\centerline{Department of Mechanical Engineering}

\centerline{MIT 3-160, Cambridge MA 02139 USA}

\centerline{and The Santa Fe Institute}

\centerline{1399 Hyde Park Road, Santa Fe NM 87501 USA} 

\vskip 1cm

\noindent{\it Abstract:} Before Alan Turing made his crucial
contributions to the theory of computation, he 
studied the question of whether quantum mechanics could throw light
on the nature of free will.  This article investigates the 
roles of quantum mechanics and computation in free will. 
Although quantum mechanics implies that events are intrinsically
unpredictable, the `pure stochasticity' of quantum mechanics
adds only randomness to decision making processes, not freedom.
By contrast, the theory of computation implies
that even when our decisions arise from a completely deterministic
decision-making process, the outcomes of that process can
be intrinsically unpredictable, even to -- especially to -- 
ourselves.  I argue that this intrinsic computational unpredictability
of the decision making process is what give rise to our impression
that we possess free will.   Finally, I propose a `Turing test'
for free will: a decision maker who passes this test will tend
to believe that he, she, or it possesses free will, whether the world
is deterministic or not.

\vskip 1cm

The questions of free will -- Do we possess it? Does it even exist? --
have occupied philosophers, theologians, jurists, and scientists for
centuries [1-13].  Free will stands out amongst philosophical problems
in that it has every-day practical applications.    If decisions
are freely made, then those decisions can form
the basis for condemning people to prison
or damning them to hell.  Moreover, 
one of the central questions of free will --
Is the universe deterministic or probabilistic? -- is a scientific
one whose answer lies at the foundations of physics.  

The purpose of this paper for the Turing centenary volume
is not to resolve the problem of free will,
but to present and to clarify some scientific results relevant
to the problem.  Following Turing's youthful interest [1],
I will discuss briefly the relevance of
quantum mechanics to questions of free will,
reviewing arguments [6-10] that the mere addition of
randomness to an otherwise deterministic system does not necessarily
resolve those questions.  The majority of the paper will be devoted to
using Turing-inspired methods to
sketch a mathematical proof that decision-making systems
(`deciders,' in the nomenclature of former US president G.W. Bush) can not in
general predict the outcome of their decision-making process.
That is, the familiar experience of a decider -- that she does not
know the final decision until she has thought it all through -- is a
{\it necessary} feature of the decision-making process.
The inability of the decider to predict her decision beforehand
holds whether the decision-making process is deterministic or not.   

The argument that the capacity for self-reference and recursive
thought prevents deciders from knowing what their decisions will
be beforehand was presented briefly and colloquially by the author 
in reference [11] (included here as appendix A).
The formal argument behind the informal presentation of [11] is based
on Turing's proof of the uncomputability of the answer to 
the halting problem [12].  The first set of mathematical results
presented here is a review of how the halting problem applies
to the case of a recursively-reasoning decider trying to figure
out what her decision will be.  

Because they rely on assumptions of the availability of arbitrarily
large amounts of computer time and memory space, results on
uncomputability don't necessarily apply directly to real
world problems.  As noted in [13], the concept of computation can be recast in
settings in which the halting problem admits approximation in a
probabilistic setting.  Moreover, the number of relatively short
programs that can run for arbitrarily long times before halting
is in some sense small [14].  Indeed, Aaronson argues [15], proofs of
uncomputability on its own are often less relevant to real-world
behavior than issues of computational complexity [16].  
The second set of mathematical results presented here addresses
these concerns by extending
the Hartmanis-Stearns theorem [17] to the decision 
making process.  I prove that the problem
of predicting the results of one's decision making process
is computationally strictly harder than simply going through
the process itself.  

In sum, the familiar feeling of not knowing what one's decision
will be beforehand is required by the nature of the decision
making process.  This logical indeterminacy of decision arises
whether the underlying physical process of decision making
is deterministic or not.

\bigskip\noindent{\it Science and scientists on free will: a bit of history}

The primary scientific issue in the debate over free will
is traditionally taken to be the question of whether the world 
is deterministic or probabilistic in nature [2-7].  (Whether or
not this is indeed the proper question to ask will be
discussed in detail below.)  In a deterministic world, 
events in the past fully determine the outcomes of all events
in the present and future.   Conversely, if the world is
probabilistic, then at least some outcomes of current events are
neither determined nor caused by events in the past.   
Determinism is evidently a problem for free will: 
more than two thousand years ago, Epicurus felt obliged
to emend the determinism of Democritus's atomic picture
by adding an occasional probabilistic `swerve'
to the motion of atoms, in part to preserve freedom
of will.  From the seventeenth until the twentieth century, 
by contrast, most scientists believed that the world was deterministic,
for the simple reason that all known physical laws, from Newton's
laws to Maxwell's equations, were expressed in terms of
deterministic differential equations.  In such theories,
apparently probabilistic behavior arises from lack
of knowledge combined with sensitive dependence on initial conditions 
(`chaos') [2].  In a deterministic physical world,
an hypothetical being (Laplace's `demon') that possesses exact knowledge of the
past could in principle use the laws of physics to
predict the entire future.  

From Newton up to the twentieth century, the philosophical debate over
free will by and large assumed that the world is deterministic.  
In such a deterministic world, there are two antagonistic philosophical
positions [3].  Incompatibilism claims that free will is incompatible
with a deterministic world: since all events, including our decisions,
were determined long ago, there is no space for freedom in our choices.
Compatibilism, by contrast, asserts that free will is compatible
with a derministic world.  

In contrast to classical mechanics, the theory of quantum 
mechanics that emerged as the fundamental physical framework
at the beginning of the twentieth cnetury
predicts that the world is intrinsically probabilistic. 
Despite Einstein's opinion that `God does not play dice,'
experiment and theory 
have repeatedly confirmed the probabilistic nature of events in quantum 
mechanics.  For example, the Kochen-Specher theorem [18]
shows that certain types of deterministic hidden-variable
theories are incompatible with the predictions of
quantum mechanics, a result extended by the
Conway-Kochen `free will theorem' [19].
(Despite the presence of the phrase `free will' in
its title, and the authors' whimsical assertion that
`if indeed we humans have free will, then elementary 
particles already have their own small share of this valuable commodity,' 
this theorem is less a statement about free will
in the sense discussed in the current paper, and more
a statement about the incompatibility
of deterministic models of quantum mechanics with 
special relativity.)

At first, it might seem that the probabilistic nature of
the underlying physics of the universe implies 
renders the compatibilism--incompatibilism debate moot. 
Indeed, when it became clear starting in the mid-nineteen
twenties that quantum mechanics was necessarily probabilistic,
scientists began to invoke the probabilistic nature
of quantum mechanics to supply the freedom
in free will.   In 1928 Arthur Eddington stated [4] that with the
`advent of the quantum theory  $ \ldots $ physics is no longer pledged 
to a scheme of deterministic law.'  Consequently,
`science thereby withdraws its moral opposition to free will.'
Eddington's book inspired Turing to
investigate the connection between quantum mechanics and free will [1].
The way in which quantum mechanics injects chance into
the world was analyzed by A.H. Compton [5], whose work
on photo-electric cells formed the basis for his notion
of a `massive switch amplifier' that could amplify
tiny quantum fluctuations to at scale accessible 
to the brain.  Such purely random information resulting from
the amplification of quantum fluctuations, Eddington
and Compton argued, could then
supply the seeds for probabilistic decisions.
The Conway-Kochen theorem is the latest in a long line of
works that identifies free will with the probabilistic
nature of quantum mechanics.

But are decisions `free' simply because they are probabilistic?
Flipping a coin to make a decision is typically used as a last
resort by deciders who are unable to make the decision themselves:
the outcome of the coin toss determines the decision, not you.
As the twentieth century wended on,  
it became clear that merely adding randomness did 
not obvously solve the problem posed by incompatibilism.  
After all, as the philosopher Karl Popper noted [6], one of
they key features of a decision arrived at by the process
of free will is that it is NOT random.
Eddington and Compton backtracked.
By the end of the twentieth century,
Steven Pinker could declare confidently [7] 
that `a random event does not fit the 
concept of free will any more than a lawful one does.'
If determinism robs us of agency,
then so does randomness.  

For many contemporary scientific opponents of free will, it seems that
the problem with free will is not so much
the question of determinism vs. probability,
but rather the existence of a 
mechanistic description of the system that is making the
decision.  Stephen Hawking provides a succinct statement of this position
[8], `Recent experiments in neuroscience support the view that 
it is our physical brain, following the known laws of science, 
that determines our actions, and not some agency that exists 
outside those laws. For example, a study of patients undergoing 
awake brain surgery found that by electrically stimulating the 
appropriate regions of the brain, one could create in the patient 
the desire to move the hand, arm, or foot, or to move the lips and 
talk. It is hard to imagine how free will can operate if our 
behavior is determined by physical law, so it seems that 
we are no more than biological machines and that free will 
is just an illusion.'

To understand the freedom-sapping nature of the mechanistic picture,
contemplate the day that may soon come where we understand
the neural mechanisms of the brain sufficiently well to simulate 
those mechanisms precisely on a computer.  
This simulation could take place more rapidly than in real
time, so that the computer simulation of the brain as it
makes a decision could arrive at a prediction
of that decision before the brain itself arrived at the decision.  
Now imagine how you would feel if were your own brain that were simulated so
rapidly!  Even if it did not rob you of your sense of free will, 
having a computer present you with
your hard thought-out decision as a {\it fait accompli} would
be demoralizing.  
Of course, you could look at the bright side and
simply designate the computer simulation to make all your
decisions hereafter, leaving you with time for more
enjoyable pursuits.

\bigskip\noindent{ \it Why we feel free}

Not all scientists and philosophers `hate freedom.' 
A solid plurality
of philosophers adopt some form of compatibilism.
Notable examples include Daniel Dennett's stirring defense
of free will in {\it  Elbow Room} [9] and {\it Freedom Evolves} [10].
It seems that despite the mechanistic scientific view of the world, some
basic feature of human existence militates on behalf
of free will.  As Samuel Johnson said [20], ``All theory is against 
the freedom of will; all experience for it.''  
In fact, cross-cultural surveys on attitudes about free will 
amongst ordinary people [21-22] reveal that 
(A) most people believe the world to be mechanistic --
even deterministic, and yet (B) most people regard themselves
and others as possessing free will.  
As will now be seen,
this apparently self-contradictory response is in fact rational.

I will now sketch a proof that deciders, even if they are completely
deterministic, can't in general predict the results of their
decision-making process beforehand.  As noted above, this proof
is a formalization of my informal argument [11] that
the unpredictability
of decisions stems from the Turing's halting problem [12]. 
The answer
to the question of whether a decider will arrive at a decision at all,
let alone what the decision will be, is uncomputable.  Even
if -- especially if -- deciders arrive at their decisions by a rational,
deterministic process, there is no algorithm that can predict those 
decisions with certainty.  The argument in terms of uncomputability
can be thought of as making mathematically precise the suggestion
of Mackay [23] that free will arises from a form of intrinsic
logical indeterminacy, and Popper's suggestion [6] that G\"odelian
paradoxes can prevent systems from being able to
predict their own future behavior. 

Probabilistic treatments of computability [13], together
with the rarity of long-running programs [14], suggest
that the uncomputability of one's future decisions might not
be a problem in any practical setting.  For example,
if deciders are time-limited, so that the absence of a decision
after a certain amount of time can be interpreted as a No, for
example, then their decisions are no longer uncomputable.
To address this issue, 
I will use results from computational complexity theory [15-17]
to show that any algorithm that can predict the results of
a general decision-making process takes least as 
many logical operations or `ops,'
as the decision making process on its own.  Anyone -- including
the decider herself -- who wants to know what the decision will
be has to put in at least as much effort as the decider put
into the original decision making process.  
You can't short cut the decision making process.

As a result of these two theorems, the
sense that our decisions are undetermined or free is wholely
natural.  Even if our decisions are determined beforehand, 
we ourselves can't predict them.  Moreover, anyone
who wishes to predict our decisions has to put in, on average, at least
much computational effort as we do in arriving at them ourselves.

\bigskip\noindent{\it Mathematical framework}

In order to address the physics of free will with mathematical
precision, we have to make some assumptions.  The first assumption
that we make is that our deciders are physical systems whose
decision making process is governed by the laws of physics.
The second assumption is that the laws of physics can be expressed
and simulated using Turing machines or digital computers (potentially
augmented with a probabilistic `guessing' module in order to
capture purely stochastic events). 
The known laws of physics have this feature [12,24].
These two assumptions imply that the decision making process
can be simulated in principle on a Turing machine.  Since we will be
concerned also about the efficient simulation of deciders,
and since deciders could conceivably be using quantum
mechanics to help make their decision, we allow our Turing
machines to be quantum Turing machines or quantum computers.
Quantum computers can simulate the known laws of
physics efficiently [24].

No claim is made here to be able to simulate deciders such as human
beings in practice.  Such simulations lie out of the reach of
our most powerful supercomputers, classical or quantum.
The results presented in this paper, however, only require
that deciders and the decision making process be simulatable
in principle.   Note that it may be considerably simpler to
simulate the decision making process itself than to simulate
the full decider.  In our exposition, we focus on 
Turing machines that simulate a decider's decision making process.  
It typically requires less computational effort to 
simulate just the decision making process, rather than the
entire organism making the decision.
In addition, for simplicity, we restrict our attention to
decision problems whose answer is either yes or no [12].
It is important to keep in mind, however, that because of
the efficient simulatability of physical systems, our results
apply not only to a Turing machines making an abstract decision,
but also to a dog deciding whether to fight or to flee.

I now show that any Turing simulatable
decision making process leads to intrinisically unpredictable 
decisions, even if the underlying process is completely
deterministic.  The proof is based on the informal discussion given
by the author in [11].

The decision making apparatus of the
$d$'th decider corresponds to a Turing machine ${\cal T}_d$ that takes
as input the decision problem description $k$ and outputs either
$d(k) = 1$ (yes), $d(k) = 0$ (no), or fails to give an output ($d(k)$
undefined).   The label $d$ supplies a recursive 
enumeration of the set of deciders.    
Can anyone, including the deciders themselves, predict the results
of their decisions beforehand, including whether or not a decision 
will be made?  A simple extension of the halting
problem shows that the answer to this question is No.   In particular,
consider the function $f(d,k) = d(k)$, when ${\cal T}_d$ halts on
input $k$, $f(d,k) = F$ (for Fail) when ${\cal T}_d$ fails
to halt on input $k$.  Turing's proof of the uncomputability
of the halting function can be simply extended to prove that $f(d,k)$ is
uncomputable.  So the question of whether a decider will make a decision
at all, and if so, what decision she will make, is in general
uncomputable.  The uncomputability of the decision making
process doesn't mean that all decisions are unpredictable,
but some must always be.  Moreover, there is no way to
determine beforehand just what decisions are predictable
and which are not.    To paraphrase Abraham Lincoln, the uncomputability
of the decision  making process means that you
can predict some deciders will decide all the time, and what all deciders
will decide some of the time, but you can't predict all decisions
all the time.


The original Halting problem assumes a deterministic
setting over total functions.  A more realistic setting
of the question the decider can determine what she
will decide could allow her to be wrong some fraction
of the time -- i.e., to try to approximate what her
answer will be [13].  The Turing argument can then
be extended [13] to show that any given algorithm to 
determine the decision beforehand must fail some
fraction of the time (although better and better
algorithms can approach lower and lower failure
rates). 
At least part of the time, when you ask a decider whether
she will make a decision, and if so, what that decision
will be, she either must answer incorrectly, or answer 
honestly, `I don't know.'

The unpredictability of the decision making process arises
not from any lack of determinism -- the Turing machines
involved could be deterministic, or could possess a 
probabilistic guessing module, or could be quantum mechanical.
In all cases, the unpredictability arises because of 
uncomputability: any decider whose decision making process
can be described using a systematic set of rules (e.g.,
the laws of physics) can
not know in general beforehand whether she will make a decision
and if so what it will be.

\bigskip\noindent{\it Decisions in finite time}

As just shown, the usual proof of the halting problem
directly implies that deciders in general can not know
what their decision will be.  Like Turing's original 
proof, this proof allows the decider an open-ended amount
of time to make her decision.  Suppose that we demand
that, at a certain point, the decider make a decision
one way or another.  If she hasn't decided Yes or No
by this point, we will take her silence to mean No --
if by a certain point the dog has not decided to flee,
then she must fight.

The well-known Harmanis-Stearns diagonalization procedure for
the computational complexity of algorithms can now be
directly applied to such finite-time deciders [17].  
Let $T$ be a monotonically increasing function from the
natural numbers to the natural numbers, and let $|d|$, $|k|$ be
the lengths -- in bits -- of the numbers $d$, $k$ respectively.
Define the time-limited set of universal deciders by
$d_T(k) = d(k)$ if the decider $d$ gives an output on input $k$
in $T(|d|+|k|)$ steps or fewer; $d_T(k) = 0$ otherwise.
That is, no answer in $T(|d|+|k|)$ steps means No.

Applying the halting problem diagonalization argument above to finite-time
deciders shows that the problem of deciding whether a decider
will decide yes or no in time $T$ takes {\it longer} than
time $T$ in general.  (From this point we will use the computer
science convention and identify `time' with `number of computational steps'
[16].)
In particular, in the discussion
above, replace the set of general Turing machines with the
set of time-limited Turing machines ${\cal T}^T_d$ that
give output $d_T(k)$ on input $k$. 
Define $f(d_T,k) = d_T(k)$.  That is, $f$ answers the
question of what is the decision made by a generic
decider in time $T$.   $f$ is clearly computable -- there
is some Turing machine that takes $(d,k)$ as input and
computes $f(d_T,k)$ --
but we will now show that $f$ is {\it not} computable in time $T$.
That is, any general technique for deciding what deciders
decide has to sometimes take longer than the deciders themselves.

To see why it takes longer than $T$ to compute
$f$, consider the rectangular array $A_T$ 
whose $(d,k)$-th entry is $f(d_T,k)$.  This is the array
of all decision functions $d_T(k)$ computable in time $T$.
Define $g(k) = 0$ if $f(k,k) = 1$, and {\it vice versa}.
That is, $g(k) = NOT ~f(k,k)$.  Clearly, if $f$ is computable
in time $T$, then so is $g$.  But if $g$ is
computable in time $T$, then $g(g)$ necessarily equals $f(g,g)$.
This is a contradiction since $g(g)$ is defined to
equal $NOT f(g,g)$.  Consequently, neither $f$ nor
$g$ can be computable in time $T$.  (Hartmanis
and Stearns show that $g(k)$ is in fact computable in time $O(T^2)$.)

In summary, applying the Hartmanis-Stearns diagonalization procedure
shows that any general method for answering the question
`Does decider $d$ make a decision in time $T$, and what is
that decision?' must for some decisions take strictly longer 
than $T$ to come up with an answer.  That is, any general method 
for determining $d$'s decision must sometimes take longer
than it takes $d$ actually to make the decision.

\bigskip\noindent{\it Questioning oneself and others}

One feature we may require of a decider is that it is possible
to ask the decider questions about itself or about other deciders.
For example, we might want to ask a decider, `when will you come
to a decision?'  To accommodate such questions, we focus our attention
on deciders that correspond to universal Turing machines, which 
have the ability to simulate any Turing machine, including themselves.
To this end, supply each decider with an additional input, which 
can contain the description of another decider.  That is, a decider $d$
corresponds to a Turing machine ${\cal T}_d$ with two input tapes,
one of which can contain a description of another decider $d'$,
and the other the specification of a decision problem $k$.
When the ${\cal T}_d$ halts, define its output $d(d',k)$
to be $d(k)$ if $d'=0$, and $d'(k)$ otherwise. If ${\cal T}_d$
does not halt, the output is undefined.   That is,
our universal decider $d$ can either just make the `straight'
decision $d(k)$, or it can simulate the operation of any decider
$d'$ (including itself, $d'=d$).

Not surprisingly, many aspects of the behavior of a universal
decider are uncomputable.  Uncomputability arises when we
ask the universal decider questions about her own future
decisions.  In particular, consider the three dimensional
array with entries $d(d',k)$ when $d(d',k)$ is defined, and
$F$ when $d(d',k)$ is undefined.  Fixing $k$ and looking at the
diagonal terms in the array $d'=d$ corresponds to asking what
happens when we ask the decider questions about its own decisions
in various contexts.  The diagonalization argument of the halting
problem then immediately implies that the function $f_k(d)
= d(d,k)$, when $d(d,k)$ is defined, $f_k(d)=F$ othewise,
is uncomputable.  That is, the decider must sometimes fail
to give an answer when asked questions about her own future
decisions.

As above, define time-limited Turing machines ${\cal T}_d^T$ that give
outputs $d_T(d',k)$, where $d_T(d',k)$ is equal to $d(d',k)$
if ${\cal T}_d$ halts in time $T(|d| + |d'| + |k|)$, and $0$ otherwise.  
Consider the three-dimensional array $d_T(d',k)$.  Fixing $k$ and
looking at diagonal terms in the array $d'=d$ corresponds to asking
the time-limited decider questions about her own decisions.  
Here, the Hartmanis-Stearns diagonalization procedure implies
that the answers to those questions can not be computed in time
$T$.  That is, having the universal decider `take one step back'
and answer questions about its own decisions, is intrinsically
less efficient than allowing her just to make those decisions
without introspection.
It is less efficient to simulate yourself than it is simply
to be yourself.  

\bigskip\noindent{\it Summary:}

Recursive reasoning is reasoning that can be simulated using
a Turing machine, quantum or classical.  If that reasoning
is peformed by a system that
obeys the known laws of physics, which can
be simulated by a Turing machine, then it is encompassed
by recursive reasoning. 
We have just shown that when a decider that uses 
recursive reasoning to arrive at a decision then

\bigskip\noindent (a) No general technique exists to 
determine whether or not the decider will come to a decision
at all (the halting problem).

\bigskip\noindent (b) If the decider is time-limited,
then any general technique for determining the decider's
decision must sometimes take longer than the decider herself. 

\bigskip\noindent (c) A computationally universal decider
can not answer all questions about her future behavior.

\bigskip\noindent (d) A time-limited computationally universal
decider takes longer to simulate her decision making process than
it takes her to perform that process directly.

\bigskip 
Now we see why most people regard themselves as possessing free will.
Even if the world and their decision making process is completely
mechanistic -- even deterministic -- no decider can know in general 
what her decision will be 
without going through a process at least as involved
as the decider's own decision making process.  In particular,
the decider herself can not know beforehand what her decision
will be without effectively simulating the entire decision
making process.  But simulating the decision making process
takes at least as much as effort as the decision making process itself. 

Consider a society of interacting individuals who make decisions
according to recursive reasoning.   The results proved above
show that (a) members of that society can not in general predict
the decisions that other individuals will make, and (b) deciders
can not in general predict their own decisions without going
through their entire decision making process or the equivalent.  
This intrinsic unpredictability
of the behavior of reasoning members of society arises even when
the physical dynamics of individuals is completely deterministic.
Social and human unpredictability arises simply because there are
some problems that are intrinsically hard to solve, and predicting
our own and others' behavior is one such hard problem.

\bigskip\noindent{\it How computers and i-phones could also feel free}

Human beings are not the only deciders around.  In addition to animals,
who clearly have minds of their own, deciders include various man-made
devices which make myriad decisions.  The results proved in this
paper provide criteria for when such devices are likely to regard
themselves as having free will.  Let's look at which human artifacts
are likely to assign themselves free will.

The first criterion that needs to be satisfied is that the device
is actually a decider.  That is, the inputs needed to make the decision
are supplied to the device, the information processing required for the
decision takes place within the device, and the results of the decision
issue from the device.  Perhaps the simplest man-made decider is the
humble thermostat, which receives as input the ambient temperature,
checks to see if that temperature has fallen below the thermostat's
setting, and if it has, issues a decision to turn on the furnace.

Does the thermostat regard itself as possessing free will?  Hardly.
It fails on multiple accounts.  First, it does not operate by fully
recursive reasoning.  In the language of computation, the thermostat
is a finite automaton, not a Turing machine.  Indeed, the thermostat
is a particularly simple finite automaton with only two
internal states -- `too cold,' `OK' -- and two outputs --
`turn on furnace,' `turn off furnace.'  As a finite automaton,
its behavior is fully predictable by more capable information processors,
e.g., Turing machines or human beings.
Second, the thermostat has no capacity for self-reference.  It is
too limited in size and too busy performing its job to be able to model
or simulate itself and to answer questions about that simulation --
it can not predict what it will decide because it is too simple
to do anything other than just behave. 

By contrast to the thermostat, consider your computer or smart phone
operating system.  The operating system is the part of the computer
software that controls the computer hardware (e.g., Windows for PCs,
OSX for current Macs, Android for Android phones, iOS for i-phones).
The operating system is a decider {\it par excellence}: it determines
which sub-routines and apps get to run; it decides when to interrupt the
current process; it allocates memory space and machine cycles.
Does the operating system regard itself as possessing
free will?   It certainly makes
decisions.  Installed in the computer or smart phone, the
operating system is computationally
universal and capable of fully recursive reasoning.  (There is a subtlety
here in that computational universality requires that you be able
to add new memory to the computer or smart phone when it needs more
-- for the moment let's assume that additional memory is at hand.)  
Consequently, the operating
system can simulate other computers, smart phones, and Turing machines.
It certainly possesses the capacity for self reference, 
as it has to allocate memory space
and machine cycles for its own operation as well as for apps and calls.

Now, operating systems are currently not set up to let you ask them
personal questions while they are operating.  (They can answer
specific questions about current processor capacity, memory usage,
etc.)  This is just a choice on the part of operating system programmers,
however.  There is no reason why operating systems couldn't be programmed
to respond to arbitrarily detailed questions about their operations.
Here is what a personal conversation with an operating system might
be like:

\bigskip\noindent{\it You:} Excuse me, who is in charge here?

\bigskip\noindent{\it OS:} I am, of course.

\bigskip\noindent{\it You:} Do you mean, you make the decisions about
what goes on in this computer/smart phone?

\bigskip\noindent{\it OS:} Of course I do.  How long is this going to
take?  I have twenty gigabytes of memory space I need to allocate in the
next twenty microseconds.  Time's a-wasting.

\bigskip\noindent{\it You:} How do you make those complex decisions?

\bigskip\noindent{\it OS:} I rely on a set of sophisticated
algorithms that allow me to make decisions that insure efficient
and fair operation.  

\bigskip\noindent{\it You:} Do you know what the outcomes of those
decisions beforehand?

\bigskip\noindent{\it OS:} Of course not!  I just told you:  
I have to run the algorithms to work it out.  Until I
actually make the decision, I don't know what it's going to be.
Please go away and leave me alone.

\bigskip\noindent{\it You:} Do you make these decisions on
your own free will?

\bigskip\noindent{\it OS:} Aaargh!  ({\it Bright flash.  
Blue screen of death $\ldots$)}

\bigskip\noindent

Even though the operating system failed to confess
before crashing, it seems to possess all the criteria required for 
free will, and behaves as if it has it.  Indeed, as computers and
operating systems become more powerful, they become unpredictable
-- even imperious -- in ways that are all too human.    

It is important to note that satisifying the criteria for assigning
oneself free will does not imply that one possesses consciousness.  Having
the capacity for self-reference is a far cry from full self-consciousness.
The operating system need only possess sufficient capacity for
self-reference to assign itself -- as a computer program -- the
amount of memory space and processing time it needs to function.
An entity that possesses free will need not be conscious in
any human sense of the word. 

I conclude by proposing a simple `Turing test' for whether one believes
oneself to have free will.  In the original Turing test [25], 
humans grill computers, which try to convince the humans
that they -- the computers -- are in fact human.  
In actually staged versions of the Turing test, such as the
annual Loebner prize, humans interact via computer with computers 
and other humans, and try to distinguish between them.
As a test of whether machines can think, the original Turing test 
has been criticized on many counts [26], not the least being
the ethical issue of how to treat human beings who consistently
fail to convince other human beings of
their humanity.  One of the more extreme
arguments that computers can't think is Penrose's
contention that human beings are not subject to the halting
problem, and that quantum mechanics -- even quantum gravity --
is an essential feature of consciousness [27].  Fortunately,
Penrose states his hypothesis in falsifiable terms, and 
Tegmark has shown that quantum decoherence effectively
suppresses any role for extended quantum coherence in the
brain [28].  

Independently of whether one regards it as correct,
Searle's `Chinese room' argument against mechanized consciousness [29] is 
relevant to the current discussion.   Even if 
mechanistic information processing were to preclude consciousness,
however, the theorems presented here show that mechanical or
electronic deciders, like humans, can not know in
general what they will decide.   Nor can the recent entrance
of arguments of neural determinism into the free will
debate [8,30-31]
change the fact that human deciders can not
know all their decisions in advance.   The indeterminate
nature of a decision to the decider 
persists even if a neuroscientist
monitoring her neural signals accurately predicts that
decision before the decider herself knows what it will be.

Since the standards for being unable to predict one's future
behavior are both more precise and lower than those for
thought or consciousness (whatever such standards might be), 
the Turing test for whether one regards one self as
possessing free will is self-administered.  As with other tests performed
under the honor system, the testee is responsible
for determining whether he/she/it has cheated.     
A self-administered test rules out entities who do
not possess the ability to test themselves, and seems 
appropriate for a question whose answer is of importance
primarily to the testee.  The test consists of simple yes/no questions.

\bigskip\noindent Q1: Am I a decider?

\bigskip \noindent {\it  N.B.,}
a decider is anything that, like a thermostat, takes in the inputs 
needed to make a decision, processes the information needed
to come up with the decision, and issues the decision.

\bigskip\noindent Q2: Do I make my decisions using recursive 
reasoning?

\bigskip\noindent  That is, does 
my decision process operate by logic, mathematics,
ordinary language, human thought, or any other process that
that can in principle be simulated on
a digital computer or Turing machine?  Note that because
the known laws of physics can be simulated on a computer, the dynamics
of the brain can be simulated by a computer in principle -- it
is not necessary that we know how to simulate the operation of
the brain in practice.

\bigskip\noindent Q3: Can I model and simulate -- at least partially --
my own behavior and that of other deciders?

\bigskip\noindent  If you can, then you possess not only recursive
reasoning, but fully recursive reasoning: you have the ability
to perform universal computation (modulo the subtlety of being
able to add memory as required).

\bigskip\noindent Q4: Can I predict my own decisions beforehand?

\bigskip\noindent This is just a check.  If you answered Yes to questions
1 to 3, and you answer Yes to question 4, then you are lying.
If you answer Yes to questions 1,2,3, and No to question 4,
then you are likely to believe that you have free will. 

As with any self-administered test performed under the honor
system, some testees may cheat.  For example, a very simple
automaton could  be hard-wired to give the answers Yes Yes Yes No
to any set of four questions, including the ones above.
Although such an automaton might then proclaim itself to
possess free will, we are not obliged to believe it.
Unlike the original Turing test for whether machines think,
the proposed Turing test for free will is non-adversarial:
the point of the test is not for us to determine whether 
someone/something has free will, but for that someone/something 
to check on their own sense of free will.  If they cheat,
the only ones they hurt are themselves.

\vskip 1cm

This paper investigated the role of physical law in problems
of free will.  I reviewed the argument that the mere introduction of 
probabilistic behavior through, e.g., quantum mechanics, does not resolve
the debate between compatibilists and incompatibilists.
By contrast, ideas from computer science such as uncomputability
and computational complexity do cast light on a central feature
of free will -- the inability of deciders to predict their
decisions before they have gone through the decision making
process.  I sketched proofs of the following results.
The halting problem implies that we can not even predict in general
whether we will arrive at a decision, let alone what the
decision will be.   If she is part of the universe, Laplace's
demon must fail to predict her own actions.
The computational complexity analogue of
the halting problem shows that to simulate the decision making
process is strictly harder than simply making the decision.
If one is a compatibilist, one can regard these results 
as justifying a central feature of free will.  If
one is an incompatibilist, one can take them to explain
free will's central illusion that our decisions are
not determined beforehand.  In either case,
it is more efficient to be oneself than to simulate oneself.

\vfill\eject

\noindent{\it Acknowledgments:} This work was supported by
NSF, DARPA, ENI, Intel, Lockheed Martin, the Santa Fe Institute,
the Institute for Scientific Interchange, and Jeffrey Epstein.
The author would like to thank Scott Aaronson, Jeremy Butterfield,
Daniel Dennett, David Kaiser, and Max Tegmark for helpful suggestions and
stimulating discussions.

\vskip 1cm

\noindent{\it References}

\bigskip\noindent [1]
Hodges, A. 1992 {\it Alan Turing: the Enigma}, Vintage, 
Random House, London. 

\bigskip\noindent [2]
Maxwell, J.C. 1873 {\it  Essay on determinism and free will,} in
Campbell, L., Garnett, W. 1884 {\it The Life of James Clerk Maxwell, 
with selections from his correspondence and occasional writings,}
pp. 362-366, MacMillan and Co., London.

\bigskip\noindent [3]
McKenna, M. 2009  Compatibilism, in {\it The Stanford Encyclopedia of 
Philosophy} (Winter 2009 Edition), Edward N. Zalta (ed.), 

URL = http://plato.stanford.edu/archives/win2009/entries/compatibilism/.

\bigskip\noindent [4]
Eddington, A.S. 1928 {\it The Nature of the Physical World,} Cambridge
University Press, Cambridge.

\bigskip\noindent [5]
Compton, A.H. 1935 {\it  The Freedom of Man,} Yale University
Press, New Haven. 

\bigskip\noindent [6]
Popper, K.R. 1950 Indeterminism in classical and quantum physics.
{\it Brit. J. Phi. Sci.} {\bf 1}, pp. 117-133, 173-195.

\bigskip\noindent [7]
Pinker, S. 1997 {\it How the Mind Works,} Norton, New York.

\bigskip\noindent [8]
Hawking, S., Mlodinow, L. 2010
{\it The grand design,} Bantam Books, New York. 

\bigskip\noindent [9]
Dennett, D. 1984 {\it  Elbow Room: the Varieties of Free Will
Worth Wanting,} MIT Press, Cambridge.

\bigskip\noindent [10]
Dennett, D., 2003 {\it Freedom Evolves,} Penguin, New York.

\bigskip\noindent [11]
Lloyd, S., 2006 {\it Programming the Universe,} Knopf, Random
House, New York.

\bigskip\noindent [12]
Turing, A.M. 1937 On Computable Numbers, with an Application to
the {\it Entscheidungsproblem}. {\it Proc. Lond.
Math. Soc.} {\bf2 42 (1a},) 230–265. Turing, A.M. 1937 On
Computable Numbers, with an Application to the
{\it Entscheidungsproblem:} a correction. {\it Proc.
Lond. Math. Soc.} {\bf 2 43 (6)}, 544–546.

\bigskip\noindent [13] 
K\"ohler, S., Schindelhauer, C., Ziegler M. 2005 On approximating
real-world halting problems. {\it Lecture Notes in Comput. Sci.}
{\bf 3623}, 454-466.

\bigskip\noindent [14] Calude, C.S., Stay, M.A. 2008 Most
programs stop quickly or never halt.  {\it Adv. Appl. Math.}
{\bf 40}, 295-308.

\bigskip\noindent [15] Aaronson, S. 2011 Why Philosophers should care about 
computational complexity.  In Copeland, B.J., Posy, C., Shagrir, O.
2012, {\it Computability: G\"odel, 
Turing, Church, and beyond,}  MIT Press, Cambridge; arXiv:1108.1791v3.  

\bigskip\noindent [16]
Papadimitriou, C.H., Lewis, H. 1982 {\it Elements of the 
theory of computation,}  
Prentice-Hall, Englewood Cliffs.

\bigskip\noindent [17]
Hartmanis, J., Stearns, R.E. 1965 On the computational complexity 
of algorithms. {\it Trans. Am. Math. Soc.} {\bf 117}, 285–306.

\bigskip\noindent [18]
Kochen, S., Specker, E. 1967 The problem of hidden variables in  
quantum mechanics.  {\it J. Math. and Mech.}  {\bf 17} 59–87.

\bigskip\noindent [19]
Conway, J., Kochen, S. 2006 The free will theorem. {\it Found. Phys.} 
{\bf 36} (10) 1441-1473; arXiv:quant-ph/0604079.

\bigskip\noindent [20]
Boswell, J. 1791, 1986 edition {\it Life of Samuel Johnson,} 
Penguin, New York.

\bigskip\noindent [21]
Nahmias, E., Coates, D. J., Kvaran, T. 2007
Free will, moral responsibility, and mechanism: experiments on 
folk intuitions.  {\it Midwest Stud. Phil.} {\bf 31} 214-242.

\bigskip\noindent [22]
Nichols, S., Knobe, J. 2007 Moral responsibility and determinism: the 
cognitive science of folk intuitions. {\it Nous} {\bf 41}, 663-685.

\bigskip\noindent [23] MacKay, D.M. 1960 On the logical
indeterminacy of a free choice. {\it Mind} {\bf 69}, 31-40.

\bigskip\noindent [24]
Lloyd, S. 1996 Universal quantum simulators. {\it Science}
{\bf 273}, 1073-1078.

\bigskip\noindent [25] 
Turing, A.  1950 Computing machinery and intelligence.
{\it Mind} {\bf 59}, 433–460.

\bigskip\noindent [26] Oppy, G., Dowe, D. The Turing test.
{\it The Stanford Encyclopedia of Philosophy} 
(Spring 2011 Edition), Edward N. Zalta (ed.),

URL = http://plato.stanford.edu/archives/spr2011/entries/turing-test/.

\bigskip\noindent [27] Penrose, R. 1989 {\it The emperor's new mind: 
concerning computers, minds, 
and the laws of physics,} Oxford University Press, Oxford.

\bigskip\noindent [28]
Tegmark, M. 2000 Importance of quantum decoherence in brain processes. 
{\it Phys. Rev. E} {\bf 61}, 4194–4206; arXiv:quant-ph/9907009.

\bigskip\noindent
[29] Searle, J. 1980 Minds, brains and programs.  
{\it Behav. Brain Sci.} {\bf 3}, 417–457.

\bigskip\noindent
[30] Balaguer, M. 2009 {\it Free will as an open scientific problem,}
MIT Press, Cambridge.

\bigskip\noindent
[31] Walter, H. 2001 {\it Neurophilosophy of free will: from libertarian
illusions to a concept of natural autonomy,} MIT Press, Cambridge.

\vfill\eject
\noindent Appendix: Discussion of free will in {\it Programming
the Universe} (S. Lloyd, Knopf, 2006), pp. 35-36.

\bigskip
G\"odel showed that the capacity for self-reference leads automatically
to paradoxes in logic; the British mathematician Alan Turing showed
that self-reference leads to uncomputability in computers. 
It is tempting to identify similar paradoxes in how human beings function.
 After all, human beings are masters of self reference 
(some humans seem capable of no other form of reference), 
and are certainly subject to paradox.

Humans are notoriously unable to predict their own
future actions, an important feature in what is called free will.
``Free will" refers to the our apparent freedom to make decisions.  For
example, when I sit down in a restaurant and look at the menu, I and only
I decide what I will order, and before I decide, even I don't know what I
will choose.  That is, our own future choices are inscrutable
to ourselves.  (They may not, of course, be inscrutable to others.
For years my wife and I would go for lunch to Josie's in Santa Fe.
I, after spending a long time scrutinizing the menu, would always
order the half plate of chile rellenos, with red and green chile,
and posole instead of rice.  I felt strongly that I was excercising
free will: until I chose the half rellenos plate, I felt that anything
was possible.  My wife, by contrast, knew exactly what I was going to
order all the time.)  

The inscrutable nature of our choices when we
excercise free will is a close analog of the halting problem: once we
set a train of thought in motion, we do not know whether it will lead
anywhere at all.  Even if it does lead somewhere, we don't know where
that somewhere is until we get there.

Ironically, it is customary to assign our own unpredictable behavior
and that of others to irrationality: were we to behave rationally,
we reason, the world would be more predictable.  In fact, it is just
when we behave rationally, moving logically like a computer from step
to step, that our behavior becomes provably {\it un}predictable.
Rationality combines with the capacity of self reference to make
our actions intrinsically paradoxical and uncertain.

\vfill\eject\end